\documentclass[aps,pre,twocolumn,superscriptaddress,nofootinbib]{revtex4-1}
\usepackage{graphicx}
\usepackage{dcolumn}
\usepackage{bm}
\usepackage{amsmath} 
\usepackage{lipsum} 
\usepackage{hyperref}
\hypersetup{colorlinks=true,linkcolor=blue,citecolor=blue,urlcolor=blue}
\usepackage[normalem]{ulem}
\usepackage{times}

\begin{document}

\title {Resonance induced by higher-order coupling diversity}

\author{Cong Liu}
\affiliation{Lanzhou Center for Theoretical Physics and Key Laboratory of Theoretical Physics of Gansu Province,
Lanzhou University, Lanzhou, Gansu 730000, China}
\affiliation{Institute of Computational Physics and Complex Systems,
Lanzhou University, Lanzhou, Gansu 730000, China}

\author{Chong-Yang Wang}
\affiliation{Institute of Computational Physics and Complex Systems,
Lanzhou University, Lanzhou, Gansu 730000, China}
\affiliation{Yangtze Delta Region Institute of University of Electronic Science and Technology of China, Huzhou, Zhejiang, 313000, China}

\author{Zhi-Xi Wu}
\email{wuzhx@lzu.edu.cn}
\affiliation{Lanzhou Center for Theoretical Physics and Key Laboratory of Theoretical Physics of Gansu Province,
Lanzhou University, Lanzhou, Gansu 730000, China}
\affiliation{Institute of Computational Physics and Complex Systems,
Lanzhou University, Lanzhou, Gansu 730000, China}

\author{Jian-Yue Guan}
\affiliation{Lanzhou Center for Theoretical Physics and Key Laboratory of Theoretical Physics of Gansu Province,
Lanzhou University, Lanzhou, Gansu 730000, China}
\affiliation{Institute of Computational Physics and Complex Systems,
Lanzhou University, Lanzhou, Gansu 730000, China}

\date{\today}

\begin{abstract}
The studies of collective oscillations induced by higher-order interactions point out the necessity of group effect in coupling modelization. As yet the related advances are mainly concentrated on nonlinear coupling patterns and cannot be straightforwardly extended to the linear ones. In present work, we introduce the standard deviation of dynamic behavior for the interacting group to complement the higher-order effect that beyond pairwise in diffusive coupling. By doing so, the higher-order effect can be flexibly extended to the linearly coupled system. We leverage this modelization to embrace the influence of heterogeneous higher-order coupling, including promoting and inhibiting effects, on the signal response for two conventional models, the globally coupled overdamped bistable oscillators and excitable FitzHugh-Nagumo neurons. Particularly, we numerically and analytically reveal that the optimal signal response can be obtained by an intermediate degree of higher-order coupling diversity for both systems. This resonant signal response stems from the competition between dispersion and aggregation induced by heterogeneous higher-order and positive pairwise couplings, respectively. Our results contribute to a better understanding of the signal propagation in linearly coupled systems.
\end{abstract}

\maketitle
\section{INTRODUCTION}

Resonance-like collective oscillation has been observed in a variety of linearly coupled systems, with examples ranging from climate change~\cite{Benzi1981,Nicolis1981}, bistable nanomechanical oscillators~\cite{Badzey2005} to sensor neurons~\cite{Hanggi2002,Lindner2004}. How to utilize resonant oscillation phenomenon to amplify signal response is of a particular interests in statistical physics and computational neuroscience~\cite{Zhang2019,Tonjes2021,Baspinar2021}. One of the most prominent examples is stochastic resonance (SR), which refers to the phenomenon that the signal response of nonlinear system can be significantly enhanced rather than blunted by an intermediate level of noise intensity~\cite{Gammaitoni1998,Faisal2008}. SR was first discussed in the context of overdamped bistable system, and then was extensively studied in excitable FitzHugh-Nagumo (FHN) neurons~\cite{Longtin1993,Volkov2003,Lindner2004,Zhu2016}. Recently, it was found that the SR effect in networked oscillators (or neurons) can be further enhanced by the specific characteristics of their interacting topology~\cite{Acebrn2007}. In particular, the long-range links of small-world networks or the hubs of scale-free networks can significantly promote the resonant response~\cite{Gao2001,Perc2007,Liu2008}.

Together with the great progress in the study of SR, another paradigmatic signal amplification mechanism is proposed as diversity-induced resonance, where an optimal collective response of globally coupled heterogeneous overdamped bistable oscillators or excitable FHN neurons can be obtained in the case of an intermediate degree of diversity~\cite{Tessone2006}. Unlike noise, diversity characterizes the static randomness or quenched disorder among the elements~\cite{Rauch2004,Hong2005}, which may stem from the intrinsic properties of entities~\cite{Perez2010}, the heterogeneity of coupling strength~\cite{Martins2010,Liu2019}, the nonuniform interaction network patterns~\cite{Chen2007,Chen2008,Zhou2011,Liu2008} and even the mismatches of the amplitudes or phase lags of the external input signals~\cite{Gosak2011,Liang2010}. For instance, the heterogeneous networks can generate a stronger resonant signal response as compared to the homogeneous ones~\cite{Liu2008}. In contrast to the weak signal response in the case of uniform couplings, the heterogeneous coupling strength can induce a bell-shaped signal amplification~\cite{Liu2019}. The surface-feeding fishes can exploit the heterogeneous phase lags between the distributed lateral line organs to determinate accurately the prey angle~\cite{Liang2010}. Given the ubiquity of noise and diversity, resonance induced by randomness or heterogeneity has been extensively investigated in all kinds of disciplines~\cite{McNamara1988,Karabalin2011,Lindner1995,Zhou2001,Zhu2021}, and the study of resonance-like behavior is an expanding field of research~\cite{XLiang2020,Scialla2021,Liang2021}.

Previous studies are mainly concentrated on the pairwise coupling of the oscillators (or neurons). Nevertheless, the interactions in empirical systems may often involve in groups of entities (three or more). For instance, recent studies reveal that higher-order interactions are prevalent in plants and predator-prey ecosystems~\cite{Levine2017,Grilli2017}, social relationship networks~\cite{Iacopini2019}, neural systems~\cite{Tlaie2019} and complicatedly coupled nonlinear oscillators systems~\cite{Gambuzza2021}. Consideration of purely linear coupling may led to the wrong predictions because of the drawback that it cannot delineate the high-order effects. It is worth pointing out that compared to the purely pairwise coupling, numerous distinct phenomena may emerge when the mixture of higher-order and pairwise couplings are considered~\cite{Battiston2020}. Usually, there are two types of impacts that the higher-order interactions can induce, either promoting (attractive) or inhibiting (repulsive)~\cite{Kovalenko2021}. It is revealed that the purely promoting higher-order interaction can induce a discontinuous phase transition in phase oscillators~\cite{Skardal2020} and social contagion~\cite{Iacopini2019}. The stable complete synchronization can be obtained by the mixture of homogeneous positive higher-order and pairwise interactions in chaotic oscillators or neural systems~\cite{Gambuzza2021}. The emergence of synchronization can even be realized by the mixture of purely negative higher-order and pairwise couplings~\cite{Kovalenko2021}. However, the impact of heterogeneous higher-order interactions (including both promoting and inhibiting effects) on resonance-like behavior is still missing. The reason may be due to the fact that the modelizations of higher-order interactions are restricted to nonlinear formulas and can not be straightforwardly extended to the linear ones, since the arbitrary summation of the pairwise interactions for linear coupling is still dyadic~\cite{Neuhauser2020}. Thus, how to build a bridge leveraging the influence of higher-order interaction and linear coupling modelization is an urgent problem.

To address this issue, we employ in this work a new minimal nonlinear term, the standard deviation of the coupled groups that is frequently used to describe the group dynamical characteristics like the synchronization or homogeneity of networked oscillators~\cite{Gambuzza2021}, to replenish the higher-order influence in diffusive coupling. By doing so, the higher-order interactions can be naturally extended to the case of linear coupling patterns. We leverage this model to explore the signal response to weak stimulus in globally coupled overdamped bistable oscillators and excitable FHN neurons with heterogeneous higher-order interactions. Intriguingly, comparing to the weak signal response in purely pairwise coupling, the bell-shaped signal response curves are obtained due to the heterogeneous higher-order interactions in both systems. Furthermore, we show numerically and analytically that these resonant behaviors emerge in both circumstances where the increase of higher-order coupling heterogeneity is either pairwise correlated or not. We finally reveal that these resonance phenomenons are caused by the dispersing and clustering of collective oscillation, resulting from the competition between the dispersion and the convergence induced by heterogeneous higher-order coupling and positive pairwise coupling, respectively.

The remainder of this paper is organized as follows. In Sec.\ref{Sec2A}, the two resonance-like phenomena in overdamped bistable oscillators are shown numerically and analytically, in which the increase of higher-order coupling diversity are pairwise correlated or not. These two types of resonant signal response are explored in excitable FHN neurons in Sec.\ref{Sec2B}. A brief summary and discussion of our main results are given in Sec.\ref{Sec2C}.

\begin{figure}[ht]
  \centering
  \includegraphics[width=0.47\textwidth,clip=false]{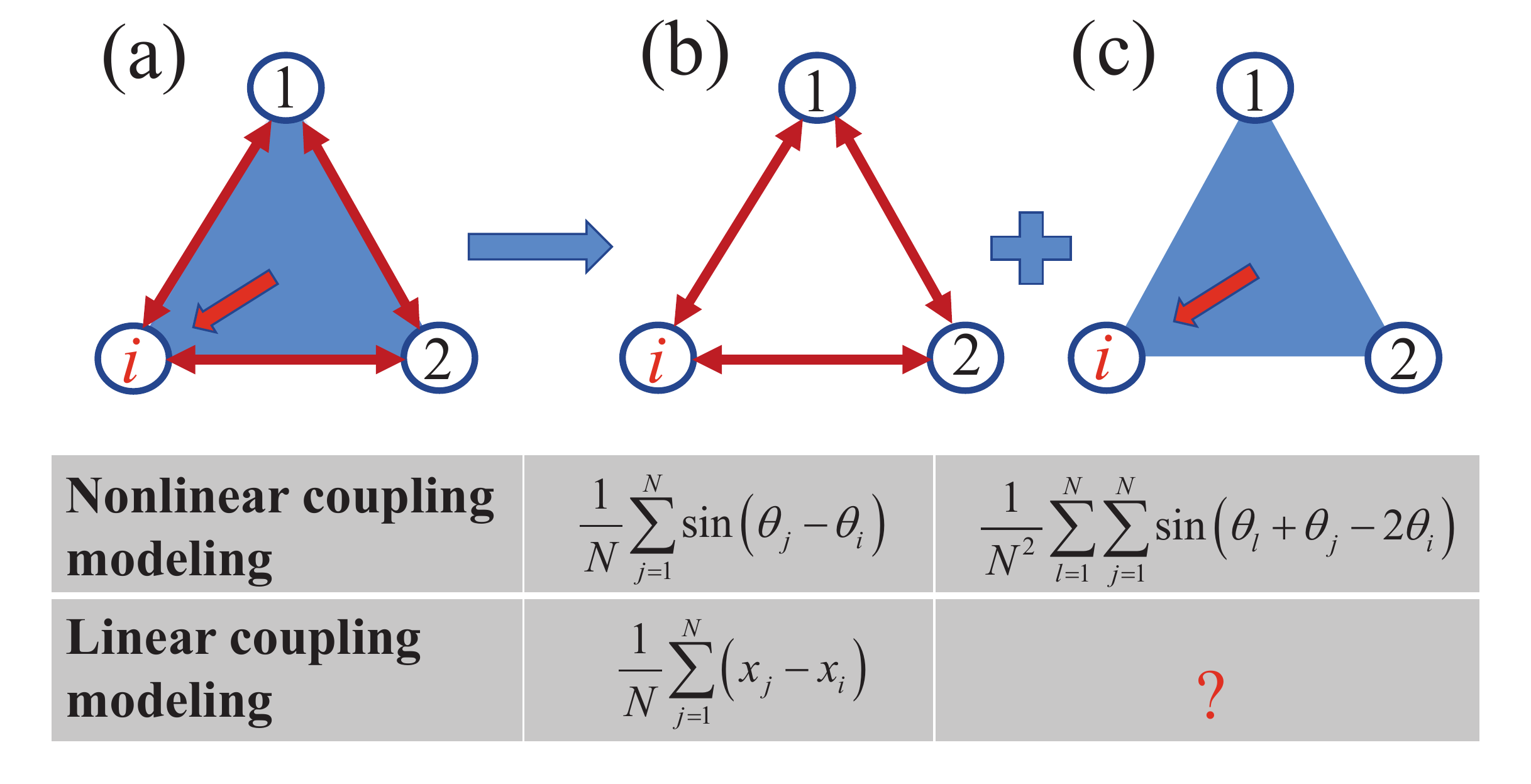}\\
  \caption{Graphical illustrations of (a) the mixture of pairwise and higher-order interactions, (b) purely pairwise interactions, and (c) purely higher-order interactions. The two conventional formulas of coupling modelings in portraying the pairwise and higher-order interactions are shown in table, respectively~\cite{Stankovski2017}. The higher-order effect can be naturally represented by the nonlinear function like sinusoidal coupling, however, the linear coupling modelings, e.g., mean-field or diffusive forms, cannot be used to interpret higher-order interactions.	}
  \label{fig 0}
\end{figure}

\section{MODEL AND ANALYSIS}

Recently, it was revealed that a single astrocyte can contact with up to $10^{5}$ synapses without anatomical connections~\cite{Tlaie2019,Pitta2016}, which suggests that a latent higher-order interaction affects efficiently the signal propagation in neuronal system. Motivated by this fact, we assume that this higher-order effect originates from a huge group, like the ensemble system, and can adaptively modulate every specific neuron~\cite{Wu2011}.

\subsection{Overdamped bistable oscillators}
\label{Sec2A}

\begin{figure*}[ht]
\centering
\includegraphics[width=1.0\textwidth,clip=false]{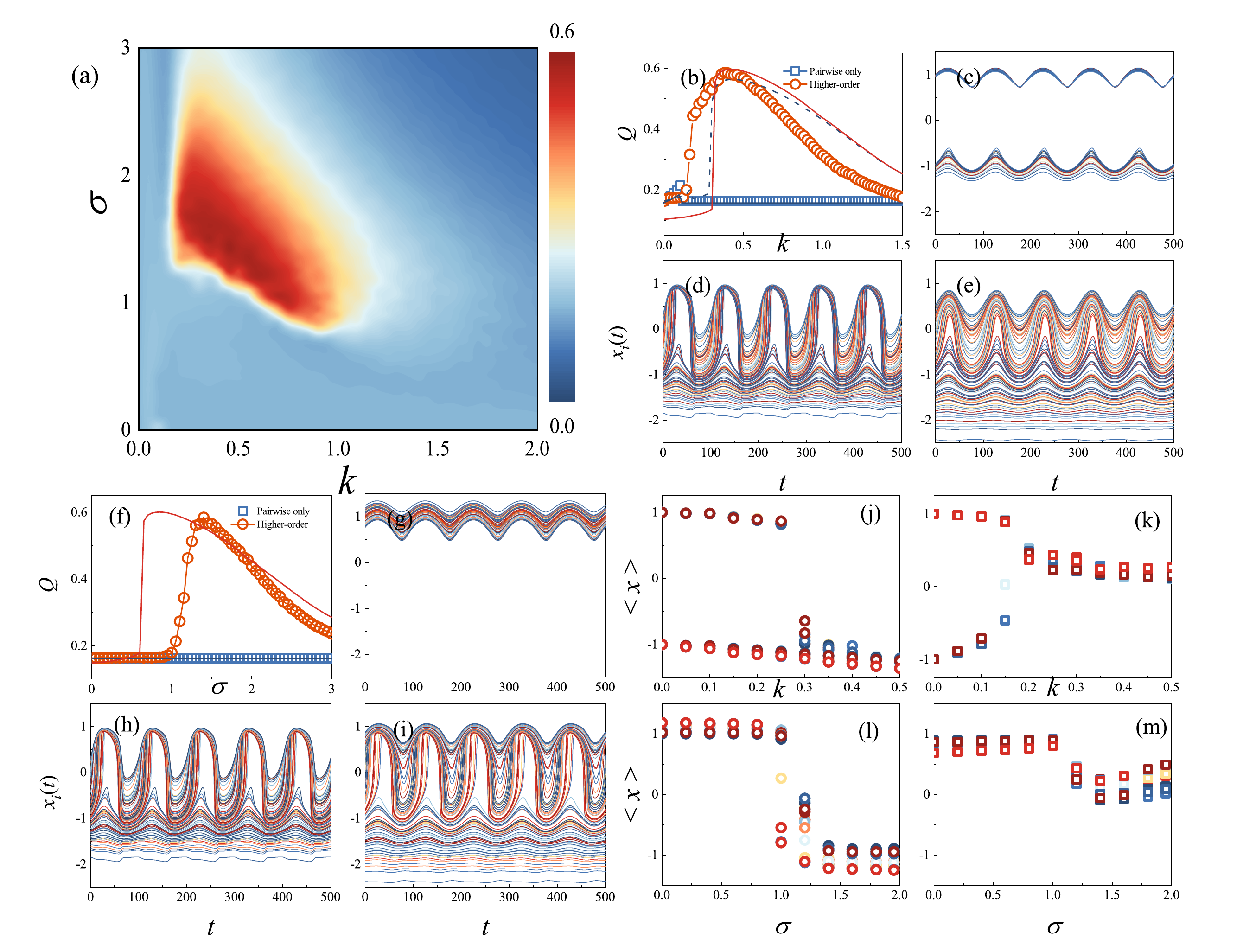}\\
 \caption{Heterogeneous higher-order coupling induced resonance in globally coupled overdamped bistable oscillators. (a) The spectral amplification factor $Q$ versus pairwise coupling $k$ and higher-order coupling diversity $\sigma$. (b) The spectral amplification factor $Q$ versus pairwise coupling $k$ for fixed higher-order coupling diversity $\sigma=1.5$. The corresponding time series $x_i(t)$ of Eq.~(\ref{eq bistable overdamped oscillators}) for three degrees of couplings are shown in (c) $k=0.1$, (d) $k=0.5$, (e) $k=1.0$, where the dynamics of $100$ nodes are selected randomly form oscillators pool. (f) The spectral amplification factor $Q$ of Eq.~(\ref{eq bistable overdamped oscillators}) versus higher-order coupling diversity $\sigma$ for fixed pairwise coupling $k=0.5$. The random selected time series $x_i(t)$ of Eq.~(\ref{eq bistable overdamped oscillators}) for three degrees of diversity are shown in (g) $\sigma=0.5$, (h) $\sigma=1.5$, (i) $\sigma=2.5$. Dependence of the fixed points $\langle x\rangle$ in different clusters on pairwise coupling $k$ or higher-order coupling diversity $\sigma$ of Eq.~(\ref{eq bistable overdamped oscillators}) are shown in (j)-(m). Diversity is fixed to $1.5$ for (j) and (k). Pairwise coupling is equal to $0.5$ for (l) and (m). The circles and squares correspond to the negative and positive higher-order-coupled entities, respectively. The higher-order couplings of ten nodes in each cluster are drawn randomly from  $(-1,0)$ and $(0,1)$ for (j), (l) and (k),(m), respectively.}
\label{Fig 1}
\end{figure*}

As bistable characteristic is prevalent in real neural and artificial signal networks, we first consider $N$ globally coupled overdamped bistable oscillators with heterogeneous higher-order coupling in response to weak signal. The dynamics of the system is described by
\begin{eqnarray}\label{eq bistable overdamped oscillators}
\dot{x}_{i}&=&x_{i}-x_{i}^{3}+\frac{k}{N}\sum\limits_{j=1}^{N}(x_{j}-x_{i})+kc_{i}(C_{v}-x_{i})\nonumber\\
&+&A\sin(\omega t),i=1,...,N,
\end{eqnarray}
where $x_{i}(t)$ denotes the state variable of the $i$th oscillator, $k$ is the uniform coupling, $C_{v}(t)=\{ \frac{1}{N}\sum_{i=1}^{N}[x_{i}(t)-X(t)]^{2} \}^{\frac{1}{2}}$ portrays the higher-order interactions, and $X(t)=N^{-1}\sum_{i=1}^{N}x_{i}(t)$ measures the collective behavior. The influence of higher-order interaction on $i$th oscillator weakens if the activity of the oscillator approaches to the average dynamic of the interacting group. This modelization reasonably captures the saturation phenomenon in neuronal activity~\cite{Gambuzza2021} and the learning behavior of agents for common opinions and interests~\cite{Wu2011}. For simplicity, the heterogeneous higher-order coupling $c_{i}$ is drawn from a Gaussian distribution with mean $\langle c_{i}\rangle=0$ and variance $\langle c_{i}c_{j}\rangle=\delta_{ij}\sigma^{2}$. The parameter $\sigma$ characterizes the level of higher-order coupling diversity. A positive coupling, $c_{i}>0$, corresponds to the oscillation aggregation induced by promoting (attractive) group influence, while a negative coupling, $c_{i}<0$, indicates oscillation dispersion caused by inhibiting (repulsive) group interaction~\cite{Liu2019}. For simplicity, we define the oscillators whose higher-order coupling strength are greater than $0$ as the positive higher-order-coupled units, and the other elements as the negative higher-order-coupled ones. $A\sin(\omega t)$ is the external weak signal with amplitude $A$ and frequency $\omega$ adding to the $i$th oscillator. Without coupling, $k=0$, each isolated oscillator can generate two distinct signal responses depending on the strength of input signal $A$. For a strong signal, $A>\sqrt{4/27}$, each oscillator undergoes a drastic oscillation around the fixed point $x^{*}=0$, while for a weak signal, $A<\sqrt{4/27}$, the oscillators jiggle slightly in one of the two potential wells depending on the initial conditions~\cite{Moss1994}. In our present work, unless specially mentioned, the system size, the amplitude and frequency of the external signal are fixed as $N=1000$, $A=0.3$ and $\omega=\pi/50$, identical to those adopted in Ref.~\cite{Liu2019}.

To quantify the signal response, the spectral amplification factor $\emph{Q}=(\emph{Q}^{2}_{\sin}+\emph{Q}^{2}_{\cos})^{\frac{1}{2}}$ is introduced~\cite{Volkov2003,Gosak2011,Liang2010}, in which 
\begin{eqnarray}\label{eq Q}
\emph{Q}_{\sin}&=&\frac{1}{nT}\int_{0}^{nT}2X(t)\sin(\omega t)dt,\nonumber\\
\emph{Q}_{\cos}&=&\frac{1}{nT}\int_{0}^{nT}2X(t)\cos(\omega t)dt,
\end{eqnarray}
where $n$ is the multiple of the period $T$. The numerical results are calculated by means of the fourth-order Runge-Kutta method with a time step $\Delta t=0.01$, $n=50$ and the initial conditions of the oscillators are chosen randomly from $x^{*}=\{-1,1\}$. The numerical results given below are ensemble averages over $200$ independent realizations.

A comprehensive view for the dependence of the signal response on both the pairwise coupling parameter $k$ and the higher-order coupling diversity parameter $\sigma$ is shown in Fig.~\ref{Fig 1}. For both the situations of identical higher-order coupling and purely pairwise coupling, the collective signal response remains weak for all degrees of coupling strength, see Figs.~\ref{Fig 1} (b) and ~\ref{Fig 1} (f). However, for the circumstance of heterogeneous higher-order coupling, one can find that an isolated red island emerges in the left center of the diagram, which implies the existence of significantly enhanced signal response for intermediate degree of pairwise coupling and higher-order coupling diversity. In detail, the two resonance-like signal response curves can be obtained by aggrandizing the pairwise coupling $k$ and higher-order coupling diversity $\sigma$, respectively, under the scheme of Eq.~\ref{eq bistable overdamped oscillators}. Noteworthy, the heterogeneity of higher-order coupling increases in two manners: (1) increasing the pairwise coupling $k$ for fixed $\sigma$, (2) increasing diversity $\sigma$ for fixed $k$. For simplicity, we define the former situation as pairwise correlated higher-order coupling diversity and the later one as pairwise uncorrelated higher-order coupling diversity.

We first fix the diversity parameter to concentrate exclusively on the impact of pairwise coupling on the resonant behavior. One can see that the signal response maintains weak until $k$ gets rise to $0.15$, then it abruptly increases to the optimal level, and gradually returns to the small one as $k$ increases from $0.4$ to $1.5$. It can be regarded as a pairwise correlated higher-order coupling diversity induced resonance, see Fig.~\ref{Fig 1} (b). The corresponding time series in Figs.~\ref{Fig 1} (c)-(e) display the oscillations of the units for different pairwise coupling strength. For a small coupling $k=0.1$, a fraction of negative higher-order-coupled oscillators are repelled to converge into the cluster that oscillates slightly around the fixed point $x^{*}=1$. The reason is attributed to the fact that the dispersion induced by the heterogeneous higher-order interactions is sufficiently high relative to the synchronization induced by the positive pairwise interactions. As a consequence, the ensemble oscillations are split into two distinct clusters, as displayed in Fig.~\ref{Fig 1} (c). For an intermediate coupling $k=0.5$, the pairwise coupling induced synchronization is strengthened so that the negative higher-order-coupled units converge into one slightly oscillating cluster. Contrariwise, the positive higher-order-coupled units oscillate sharply around $x^{*}=0$ in the other cluster, and the ensemble oscillation is consequently enhanced, see Fig.~\ref{Fig 1} (d). For a strong coupling $k=1.0$, besides the oscillations around $x^{*}=0$ are reduced, a part of units are split from the drastic oscillating cluster to a new one that oscillates slightly between $x=0$ and $x=1$, and the signal response is consequently decreased, as shown in Fig.~\ref{Fig 1} (e). A similar oscillation clustering and dispersing process induced by the pairwise uncorrelated higher-order coupling diversity can be observed in Figs.~\ref{Fig 1} (g)-(i) for fixed pairwise coupling, and the resulting bell-shaped signal response is plotted in Fig.~\ref{Fig 1} (f).

To figure out the mechanism behind the oscillation cluster, for simplicity, we define the clusters jiggling below and above $x=0$ as cluster $1$ and $2$, respectively, and the cluster oscillating between $x=1$ and $x=-1$ as cluster $3$~\cite{Liu2019,Supplemental2022}. We plot in Figs.~\ref{Fig 1} (j) and ~\ref{Fig 1} (k) the relationship between the fixed points $\langle x\rangle$ of the entities in the three oscillation clusters and the pairwise coupling parameter $k$ for a fixed diversity. One can find that the negative higher-order-coupled units show two types of transitions depending on their initial positions: (1) for $x_{i}(0)=-1$, the units oscillate slightly in cluster $1$ at the whole range of coupling parameters $k$, (2) for $x_{i}(0)=1$, the elements in cluster $2$ abruptly converge into cluster $1$ and oscillate slightly around the fixed point $x^{*}=-1$ as the coupling parameter goes beyond $0.25$. While for the positive higher-order-coupled oscillators, as $k$ increases, the upper fixed point (above $\langle x\rangle=0$) approaches to $\langle x\rangle=0$. The lower fixed point (below $\langle x\rangle=0$) gradually disappears. It behaves an imperfect (pitchfork) bifurcation~\cite{Supplemental2022,Strogatz2001}. As a consequence, the oscillations of positive higher-order-coupled units are significantly enhanced. For the situation of fixed coupling strength, e.g., $k=0.5$, see Figs.~\ref{Fig 1} (l) and (m), the fixed point of the negative higher-order-coupled units is equal to $x^{*}=1.0$ for $\sigma<1.0$, then it dispersedly translates into $x^{*}=-1$ for $\sigma$ increases. It is because that the degree of oscillation synchronization induced by the positive pairwise coupling is fixed for $k=0.5$, when $\sigma$ increases, the collective dynamics undergo a transition from oscillation synchronization to dispersion. While for the positive higher-order-coupled units, the fixed point keeps close to $x^{*}=1$ until $\sigma$ rises to $1$, subsequently, it decreases to $x^{*}=0$ for $\sigma$ increases to $1.3$, and then increases gradually as $\sigma$ increases to $2.0$.  These results suggest that a part of positive higher-order-coupled entities start to oscillate sharply as the dynamic bifurcation when the diversity parameter is slightly greater than $1.0$. For $\sigma$ approaches to $1.3$, all positive higher-order-coupled oscillators undergo dynamic bifurcation. When $\sigma$ is increasing further, a part of entities whose higher-order couplings are sufficiently large return to the slight shake.

In what follows, we give theoretical analysis on the emergent resonance phenomenon. For the situation of pairwise correlated higher-order coupling diversity, we assume that the collective oscillations are composed of two distinct oscillating clusters: the sharp oscillation $s_{1}(t)$ and the slight joggle $s_{2}(t)$  of the positive and negative higher-order-coupled entities, respectively~\cite{Supplemental2022}. The perfect synchronous oscillation in each cluster is considered for the analyses~\cite{Liang2020}. As a consequence, the collective dynamics of Eq.~(\ref{eq bistable overdamped oscillators}) can be rewritten as
\begin{eqnarray}\label{eq twogroupclusters}
\dot{s_{1}}&=&s_{1}-s_{1}^{3}+\frac{k}{2}(s_{2}-s_{1})+\frac{3k}{4}(C_{v}-s_{1})+A\sin(\omega t),\nonumber\\
\dot{s_{2}}&=&s_{2}-s_{2}^{3}+\frac{k}{2}(s_{1}-s_{2})-\frac{3k}{4}(C_{v}-s_{2})+A\sin(\omega t).
\end{eqnarray}
The signal response of Eq.~(\ref{eq twogroupclusters}) is shown by the blue dashed line in Fig.~\ref{Fig 1} (b), which captures well the resonant characteristic. Introducing $S(t)=(s_{1}+s_{2})/2$ and $M(t)=(s_{1}-s_{2})/2$ to represent the average activity and dynamics deviation of the two clusters, respectively. The ensemble dynamic can be written as
\begin{eqnarray}\label{eq Deviation behavior}
\dot{S}&=&(1-3M^{2})S-S^{3}-\frac{3}{4}kM+A\sin(\omega t),\nonumber\\
\dot{M}&=&(1-3S^{2}-k)M-M^{3}+\frac{3}{4}k(C_{v}-S).
\end{eqnarray}
Particularly, under the scheme of Eq.~(\ref{eq twogroupclusters}), the higher-order effect indicator $C_{v}$ is equal to the dynamics deviation. Equation~(\ref{eq Deviation behavior}) can be further reduced to
\begin{eqnarray}\label{eq Group behavior}
\dot{S}&=&(1-3C_{v}^{2})S-S^{3}-\frac{3}{4}kC_{v}+A\sin(\omega t),\nonumber\\
\dot{C}_{v}&=&(1-3S^{2}-\frac{1}{4}k)C_{v}-C_{v}^{3}-\frac{3}{4}kS.
\end{eqnarray}
Assuming the change of $C_{v}(t)$ is much slower than $S(t)$, we obtain the dynamic of $C_{v}(t)$ by the adiabatic elimination approach~\cite{Liang2020} as
\begin{equation}\label{eq Deviation}
C_{v}(t)=-2\sqrt{\frac{\beta}{3}}\sinh\left[\frac{1}{3}\text{arsinh}\left(\frac{2.25kS}{2\beta}\sqrt{\frac{3}{\beta}}\right)\right],
\end{equation}
where $\beta=3S^{2}+0.25k-1$. Incorporating Eq.~(\ref{eq Deviation}) to Eq.~(\ref{eq Group behavior}), we yield the reduced ensemble dynamics,
\begin{equation}\label{eq REDUCED}
\dot{S}=(1-3C_{v}^{2})S-S^{3}-\frac{3}{4}kC_{v}+A\sin(\omega t),
\end{equation}
where $C_{v}(t)=-2\sqrt{\frac{\beta}{3}}\sinh\left[\frac{1}{3}\text{arsinh}\left(\frac{2.25kS}{2\beta}\sqrt{\frac{3}{\beta}}\right)\right]$.
Inserting Eq.~(\ref{eq REDUCED}) into Eq.~(\ref{eq Q}), the semianalytical signal response is shown in Fig.~\ref{Fig 1} (b) by the red line, which capture well the main features of the dynamical behaviors of Eq.~(\ref{eq bistable overdamped oscillators}).

For the case of resonance induced by pairwise uncorrelated higher-order coupling diversity, for simplicity, one can consider the pairwise correlated higher-order coupling diversity as a specific situation of the pairwise uncorrelated diversity, in which the impact of pairwise coupling is weak and could be neglected. We yield the relationship of $\sigma*0.5=k$. Inserting it into Eq.~(\ref{eq Deviation}), the reduced higher-order behavior can be represented as
\begin{equation}\label{eq fixedcoupling}
C_{v}(t)=-2\sqrt{\frac{\beta}{3}}\sinh\left[\frac{1}{3}\text{arsinh}\left(\frac{1.5\sigma S}{2\beta}\sqrt{\frac{3}{\beta}}\right)\right],
\end{equation}
where $\beta=3S^{2}-1$. Similarly, combing Eq.~(\ref{eq Group behavior}) and Eq.~(\ref{eq fixedcoupling}), one can obtain the reduced collective dynamics and the semianalytical signal response by inserting the reduced equation into Eq.~(\ref{eq Q}). The theoretical results shown in Fig.~\ref{Fig 1} (f) by the red line captures well the resonant behaviors of Eq.~(\ref{eq bistable overdamped oscillators}).


\subsection{Excitable FHN neurons }
\label{Sec2B}

A more biologically realistic framework, the signal response of $N$ electrically coupled FHN neurons with heterogeneous higher-order coupling, is now considered to model the information interchange among neurons and astrocytes. The dynamics of the system are described by the following differential equations~\cite{Zhou2001,Scialla2021,Fitzhugh1961}
\begin{eqnarray}\label{eq FHN}
\epsilon\dot{x}_{i}&=&x_{i}(1-x_{i})(x_{i}-a)-y_{i}+\frac{k}{N}\sum\limits_{j=1}^{N}(x_{j}-x_{i})\nonumber\\
&+&kc_{i}(C_{v}-x_{i})+A\sin(\omega t),\nonumber\\
\dot{y}_{i}&=&bx_{i}-y_{i}-d,  i=1,...,N,
\end{eqnarray}
where $x_{i}$ and $y_{i}$ represent the fast membrane potential and slow potassium gating variable, respectively. The parameter $\epsilon=0.01$ characterizes the timescale separation between the fast and slow variables, $a=0.5$ controls the excitability, $b=0.1$ governs the interplay between the fast and slow variables, and the constant $d$ is fixed to $0.05$. In this formulation, an isolated FHN neuron is both excitable and bistable with two fixed points $(x^{*}_{1},y^{*}_{1})=(0.89,0.04)$ and $(x^{*}_{2},y^{*}_{2})=(0.11,-0.04)$. To model a subthreshold signal, the amplitude and frequency are fixed to $A=0.01$ and $\omega=\pi/5$, respectively, as the reference~\cite{Liu2020} suggests, so that each neuron cannot generate spikes but can only oscillate slightly around the two fixed points. The time step of simulations is $\Delta t=0.001$ and the signal response can be revealed by the spectral amplification factor $Q$ as well, in which $X(t)$ represents the collective activity of the fast membrane potential. The random parameters $c_{i}$ are drawn from the standard normal distribution as in bistable oscillators, and the initial conditions of the neurons are randomly chosen from $(x^{*}_{1},y^{*}_{1})$ and $(x^{*}_{2},y^{*}_{2})$.

\begin{figure*}[ht]
\centering
\includegraphics[width=1.0\textwidth,clip=false]{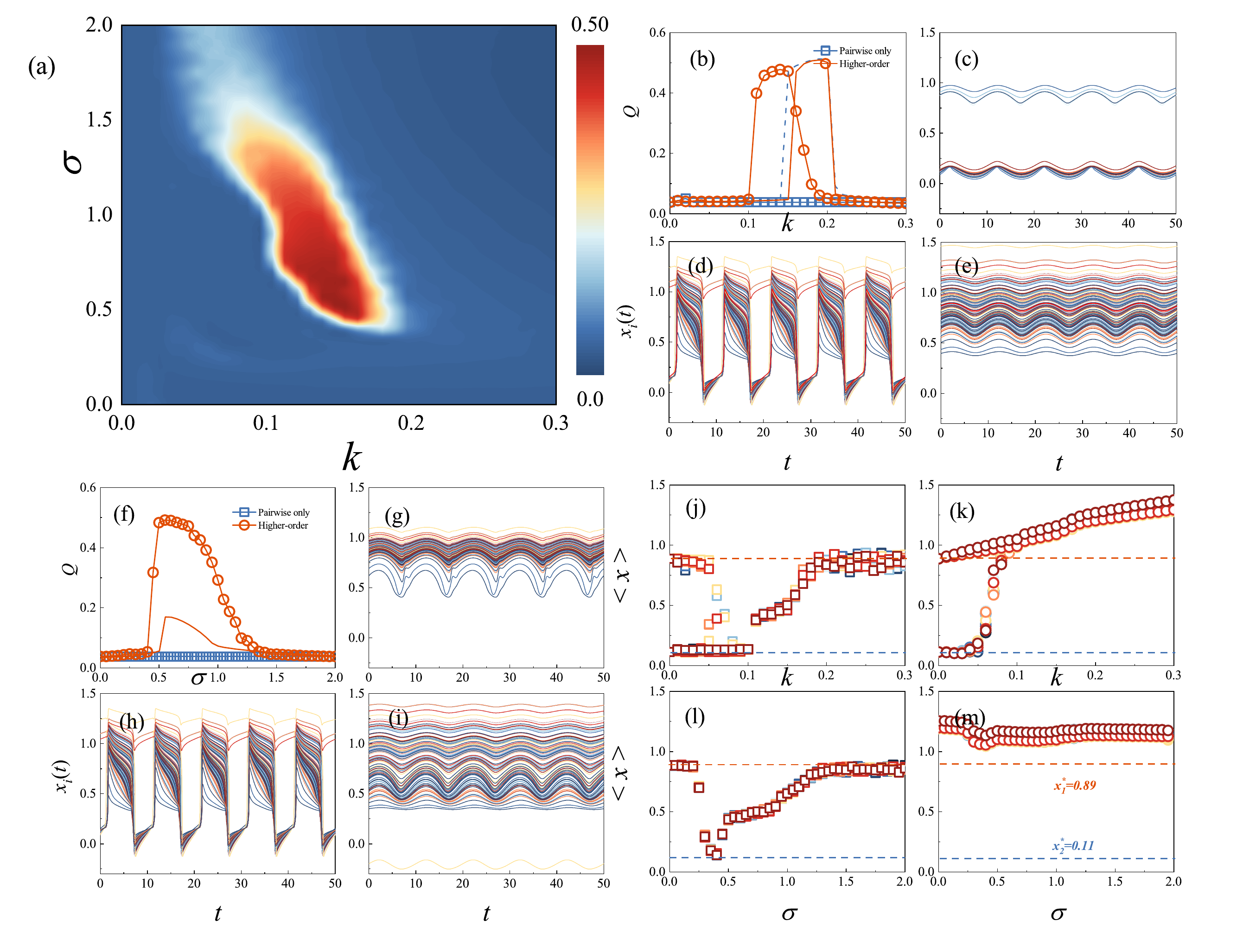}\\
 \caption{ Heterogeneous higher-order coupling induced resonance in globally coupled FHN neurons. (a) The spectral amplification factor $Q$ versus pairwise coupling $k$ and higher-order coupling diversity $\sigma$. (b) The spectral amplification factor $Q$ versus pairwise coupling $k$ for fixed higher-order coupling diversity $\sigma=0.75$. The corresponding time series $x_i(t)$ of Eq.~(\ref{eq FHN}) for three degrees of couplings are shown in (c) $k=0.05$, (d) $k=0.15$, (e) $k=0.25$, where the dynamics of $100$ neurons are selected randomly form elements pool. (f) The spectral amplification factor $Q$ of Eq.~(\ref{eq FHN}) versus higher-order coupling diversity $\sigma$ for fixed pairwise coupling $k=0.15$. The neural activities $x_i(t)$ of the random selected neurons for three degrees of diversity are shown in (g) $\sigma=0.25$, (h) $\sigma=0.75$, (i) $\sigma=1.5$. Dependence of the fixed points $\langle x\rangle$ in different clusters on pairwise coupling $k$ or higher-order coupling diversity $\sigma$ of Eq.~(\ref{eq FHN}) are shown in (j)-(m). Diversity is fixed to $0.75$ for (j) and (k). Pairwise coupling is equal to $0.15$ for (l) and (m). The squares and circles correspond to the positive and negative higher-order-coupled entities, respectively. The higher-order couplings of ten nodes in each cluster are drawn randomly from $(-1,1)$ and $(-2,-1.5)$ for (j), (l) and (k),(m), respectively.}
\label{Fig 2}
\end{figure*}

In Fig.~\ref{Fig 2} (a), in analogy to the signal amplification in bistable oscillators, we find that an prominently enhanced signal response island emerges in the center of the graph, which indicates that an intermediate degree of higher-order coupling diversity and pairwise coupling can significantly enhance the signal response. When only pairwise interactions among the neurons are taken into account, see the blue squares in Fig.~\ref{Fig 2} (b), the ensemble signal response is weak at the whole range of coupling strength. While for the mixture of heterogeneous higher-order and pairwise interactions, the signal response maintains feeble for $k<0.1$, then it grows abruptly to the maximum, and subsequently tapers off as the pairwise coupling strength increases, which demonstrates that the the pairwise correlated higher-order coupling diversity can give rise to the resonance phenomenon in excitable system. The corresponding time series uncover the resonance-like signal response for different coupling strengths. For a small coupling strength $k=0.05$, the collective dynamics split into two slightly oscillating clusters when the synchronization induced by the positive pairwise coupling is weak with respect to the dispersion induced by the fixed higher-order coupling diversity, see Fig.~\ref{Fig 2} (c). For an intermediate coupling strength $k=0.15$, the two distinct clusters (periodic spiking and small-amplitude oscillation) driven by the external force can be observed in Fig.~\ref{Fig 2} (d). Thus, the signal response is significantly enhanced. For a large coupling strength $k=0.25$, all neurons oscillate slightly because the synchronization induced by the positive pairwise coupling are strong with respect to dispersion, see Fig.~\ref{Fig 2} (e). As a consequence, the signal response is reduced. Additionally, a similar dynamic clustering and the resulting bell-shaped signal response curve can be achieved by tuning the pairwise uncorrelated higher-order coupling diversity as well, see Fig.~\ref{Fig 2} (f)-(i).

To show how does the heterogeneous higher-order coupling cause the neurons to discharge, similar to bistable system, we define the clusters oscillating below and above $x=0.25$ as cluster $1$ and $2$, respectively, and the cluster discharging between $x=0$ and $x=1$ as cluster $3$~\cite{Supplemental2022}. The dependence of the fixed points $\langle x \rangle$ in different clusters on pairwise coupling parameter $k$ or higher-order coupling diversity parameter $\sigma$ are given in Figs.~\ref{Fig 2} (j)-(m). For both positive and weak negative higher-order-coupled neurons~\cite{Supplemental2022}, one can find that the neurons in cluster $2$ gradually converge into cluster $1$, and oscillate with small amplitude around the uniform fixed point $x^{*}=0.11$ as $k$ increases to $0.1$. For $0.1<k<0.15$, the fixed point abruptly grows to $x^{*}=0.5$, which demonstrates that these neurons undergo the saddle-node bifurcation and begin to fire periodically~\cite{Gerstner2014,Supplemental2022}. By increasing $k$ further, the firing activities of the neurons in cluster $3$ are gradually reduced, and the ensemble neurons finally converge into cluster $2$ as $k$ increases to $0.2$. The arise of this phenomenon is due to the improvement of synchronization induced by the positive pairwise couplings, see Fig.~\ref{Fig 2} (j). For strong negative high-order-coupled neurons, see Fig.~\ref{Fig 2} (k), as the pairwise coupling strength increases, they gradually merge into cluster $2$ and oscillate slightly driven by the external signal. For the case of pairwise uncorrelated higher-order coupling, see Fig.~\ref{Fig 2} (l), as $k=0.15$, the synchronization induced by positive pairwise coupling is sufficiently strong so that both the positive and small negative higher-order-coupled neurons converge into cluster $2$ for $\sigma<0.2$. For $0.2<\sigma<0.4$, the neurons in cluster $2$ gradually converge to cluster $1$. For $0.4<\sigma<0.8$, the fixed point increases abruptly to $0.5$, which implies the neurons undergo dynamic bifurcation. If $\sigma$ increases further, the fixed point get gradually rise to $0.89$, which demonstrates that the firing neurons return back to the small-amplitude oscillations. However, see Fig.~\ref{Fig 2} (m), the states of the strong negative higher-order-coupled neurons keep unchanged for the whole range of the magnitude of higher-order coupling diversity.

Similar to the case of bistable system, the collective dynamics in FHN neurons can be explained as the combination of the two distinct clusters: one represents the spiking activity $s_{1}(t)$, and the other stands for the small-amplitude oscillation driven by the external signal $s_{2}(t)$~\cite{Supplemental2022}. Then, the collective behaviors of Eq.~(\ref{eq FHN}) can be formulated as
\begin{eqnarray}\label{eq two clustersFHN1}
\epsilon\dot{s}_{1}&=&s_{1}(1-s_{1})(s_{1}-a)-\mu_{1}+\frac{k}{2}(s_{2}-s_{1})\nonumber\\
&+&\frac{1}{2}k(C_{v}-s_{1})+A\sin(\omega t),\nonumber\\
\dot{\mu}_{1}&=&bs_{1}-\mu_{1}-c,\nonumber\\
\epsilon\dot{s}_{2}&=&s_{2}(1-s_{2})(s_{2}-a)-\mu_{2}+\frac{k}{2}(s_{1}-s_{2})\nonumber\\
&-&\frac{1}{2}k(C_{v}-s_{2})+A\sin(\omega t),\nonumber\\
\dot{\mu}_{2}&=&bs_{2}-\mu_{2}-c.
\end{eqnarray}

Considering $S(t)=(s_{1}+s_{2})/2$, $U(t)=(\mu_{1}+\mu_{2})/2$, $M(t)=(s_{1}-s_{2})/2$, $\Delta(t)=(\mu_{1}-\mu_{2})/2$, and $C_{v}(t)=-M(t)$, the ensemble dynamics of the system can be written as
\begin{eqnarray}\label{eq meanFHN}
\epsilon\dot{S}&=&-(0.5+3M^{2})S+1.5(S^{2}+M^{2})-S^{3}\nonumber\\
&-&U-0.5kM+A\sin(\omega t),\nonumber\\
\dot{U}&=&bS-U-c,\nonumber\\
\epsilon\dot{M}&=&(-0.5-1.5k+3S-3S^{2})M-M^{3}-\Delta-0.5kS,\nonumber\\
\dot{\Delta}&=&bM-\Delta.
\end{eqnarray}
The predictions obtained by Eq.~(\ref{eq FHN}) and Eq.~(\ref{eq meanFHN}) agree reasonably well, as shown by the blue dash line and red circles in Fig.~\ref{Fig 2} (b), which demonstrates that the two cluster reduced equations can capture essentially the behavior of the ensemble FHN neurons.

For simplicity, we assume $\dot{M}=0$ and $\dot{\Delta}=0$ by using the approximation of adiabatic elimination~\cite{Liu2020}. Then, the ensemble dynamics are read as
\begin{eqnarray}\label{eq ReducedFHN equation}
\epsilon\dot{S}&=&-(0.5+3M^{2})S+1.5(S^{2}+M^{2})-S^{3}\nonumber\\
&+&A\sin(\omega t)-U-0.5kM,\nonumber\\
\dot{U}&=&bS-U-c,
\end{eqnarray}
where $M(t)=-2\sqrt{\frac{\beta}{3}}\sinh\left[\frac{1}{3}\text{arsinh}\left(\frac{3kS}{2\beta}\sqrt{\frac{3}{\beta}}\right)\right]$, $\beta=0.6+2k-3S+3S^{2}$. The results predicted by Eq.~(\ref{eq ReducedFHN equation}) are displayed by the red line in Fig.~\ref{Fig 2} (b), which represents explicitly a resonant signal response as $k$ increases.

Considering the case of pairwise correlated higher-order coupling diversity as the specific situation of the pairwise uncorrelated one, where the impact of pairwise coupling is weak and could be neglected, we can obtain the relation $k=0.15*\sigma$. Inserting the relation and $\dot{U}=0$ into Eq.~(\ref{eq ReducedFHN equation}), the reduced relationship between the collective dynamics and the higher-order coupling diversity is read as
\begin{eqnarray}\label{eq Reducedequation}
\epsilon\dot{S}&=&-(0.5+3M^{2})S+1.5(S^{2}+M^{2})-S^{3}\nonumber\\
&-&bS-c-0.5kM+A\sin(\omega t),
\end{eqnarray}
where $M(t)=-2\sqrt{\frac{\gamma}{3}}\sinh\left[\frac{1}{3}\text{arsinh}\left(\frac{1.5kS}{2\gamma}\sqrt{\frac{3}{\gamma}}\right)\right]$, $\gamma=0.6+1.5k-3S+3S^{2}$.
Combining Eq.~(\ref{eq Reducedequation}) and Eq.~(\ref{eq Q}), we yield the semianalytical bell-shaped signal response, see the red line in Fig.~\ref{Fig 2} (f). 

\section{DISCUSSION AND CONCLUSIONS}
\label{Sec2C}

How heterogeneous higher-order interactions shape the signal response in a linear coupling pattern is a question of great relevance to many man-made and natural systems, e.g., the signal propagation in neurons and information diffusion in social systems. To address these issues, by utilizing the standard deviation that is frequently used to evaluate group synchronous status of coupled oscillators to represent the higher-order coupling of interacting group, we have investigated the signal response of globally coupled overdamped bistable oscillators and excitable FHN neurons, in which the coupling patterns are delineated by the mixture of pairwise and higher-order interactions. The pairwise couplings are homogeneous and the higher-order couplings are drawn from a normal distribution. We have found that an intermediate degree of higher-order coupling heterogeneity can significantly enhance the signal response in both systems. Particularly, the resonant phenomenon emerges in the two circumstances where the increase of higher-order coupling diversity are associated with pairwise coupling or not. We have demonstrated that the oscillating dispersion induced by heterogeneous higher-order interactions and the convergence induced by positive pairwise interactions jointly regulate the dynamic bifurcation and clusters forming. More precisely, for a small degree of diversity, all units cannot undergo dynamic bifurcation, and the signal response is weak. While for a moderate degree of diversity, on one hand, the overdamped bistable oscillators (FHN neurons) oscillate drastically (discharge periodically) resulting from the dynamic bifurcation of the positive ( both positive and weak negative) higher-order-coupled oscillators (neurons). On the other hand, the number of intensively oscillating units (discharging neurons) reaches the maximum~\cite{Supplemental2022}. As a consequence, the signal response is optimal. For a large degree of diversity, both the number and amplitude of intensively oscillating units (discharging neurons) decreases, the resonance phenomenon is accordingly reduced. Finally, since the collective dynamics are composed of the two distinct oscillations, we proposed a reduced equation to estimate the response of the system, which predicts well the numerical results.

Our findings demonstrate that the heterogeneous higher-order interactions including promoting and inhibiting effects may play an important role in signal amplification. Furthermore, the modeling of higher-order interaction in this work can also be extended to the complex networked system like random simplicial complex~\cite{Iacopini2019}. Our modelization of the higher-order interation breaks the restriction that the higher-order coupling can only be modeled in nonlinear couplings and thus could be a step toward the realistic modeling of linearly coupled systems, i.e., signal propagation in neuron systems with higher-order interactions.

\section*{ACKNOWLEDGMENTS}
C.L. thanks Xiaoming Liang and Xiyun Zhang for fruitful discussions. We acknowledge financial support from the National Natural Science Foundation of China (Grants No. 11975111 and No. 12047501).

\bibliography{Mybibfile}
\end{document}